\documentclass[twocolumn]{aa} % for a paper on 2x column  
\usepackage{graphicx}
\usepackage{txfonts}
\begin{document}
   \title{Chromaticity in all-reflective telescopes for astrometry}

   \author{D.Busonero\inst{1,2},
          M.Gai\inst{1},
	  D.Gardiol\inst{1},
          M.G.Lattanzi\inst{1},
	  D.Loreggia\inst{1}          }

          \offprints{M. Gai}

    \institute{National Institute of Astrophysics (INAF), Astronomical
Observatory of Turin,
              Via Osservatorio 20, I-10025 Pino Torinese, Turin, Italy\\
              \email{gai@to.astro.it, loreggia@to.astro.it, gardiol@to.astro.it,
lattanzi@to.astro.it}
         \and
             University of Siena, Department of Physics, Via Roma 56,
I-53100 Siena, Italy\\
             \email{busonero@to.astro.it}
             }

\date{Received 9 September 2005/ Accepted 20 October 2005}

\abstract {}{Chromatic effects are usually associated with
refractive optics, so reflective telescopes are assumed to be
free from them. We show that all-reflective optics still bears
significant levels of such perturbations, which is especially critical to
modern micro-arcsecond astrometric experiments.}{We analyze the
image formation and measurement process to derive a precise
definition of the chromatic variation of the image position, and
we evaluate the key aspects of optical design with respect to
chromaticity.}{ The fundamental requirement related to chromaticity
is the symmetry of the optical design and of the wavefront errors.
Finally, we address some optical engineering issues, such as
manufacturing and alignment, providing recommendations to minimize
the degradation that chromaticity introduces into astrometry.}
{}
   \keywords {Astrometry, Methods: data analysis, Space vehicles:
     instruments, Techniques: high angular resolution, Telescopes.}
   
\authorrunning{D.Busonero et al.}

\titlerunning{Chromaticity in all-reflective telescopes for astrometry}
   \maketitle
%
%________________________________________________________________
\section{Introduction}
Chromatism is usually defined for refractive optics as an aberration due to the
light dispersion of the glass with the refractive index; this induces a
perturbation on the image profile, depending upon the source color. In
astrometry, it provides a variation in the apparent star position, i.e. an
astrometric error.\\
This effect can be reduced by using more complex refractive systems (doublets,
triplets, etc.), taking advantage of the different dispersion of glasses to
achieve a certain degree of compensation. For most purposes, the only way to
avoid chromatism is considered to be adoption of a completely reflecting design. Here, we show that, at the demanding level of modern astrometry,
reflective optics is still affected by significant chromatic effects, which we
will refer to as "{\it chromaticity}". We investigate the source of the
chromaticity and provide recommendations for its minimization in optical
design, manufacturing, and alignment.\\
Currently, space astrometry experiments are being designed and implemented with the goal of micro-arcsecond (hereafter, $\rm{\mu as}$) measurements. We
will refer to the framework of the Gaia mission, approved within the space
science program of ESA for launch before 2012. Built upon the implementation
of the Hipparcos (\cite{Hipptycho}) concept, with the benefit of modern
technology and more advanced astrophysical understanding, Gaia aims to measure
absolute position, parallax and annual proper motion of $\sim 10^{9}$ objects
with a typical accuracy of $\sim$ 15 $\rm{\mu as}$ for V=15 stars, with survey
completeness to V=18 and limiting magnitude V=20
(\cite{Perryman2}). Hipparcos's payload was already affected by chromaticity
at the mas-level, and during Gaia's design phase it soon became apparent that
its all-reflective optics would also be affected by chromaticity at a similar
level (\cite{Lattanzi}). Therefore chromaticity in Gaia must be suppressed
by more than two orders of magnitude.\\
We describe how the aberrations of a realistic all-reflective optical
system can lead to chromaticity as a consequence of light diffraction, even
when propagation occurs in a practically non-dispersive medium as would be
the case for payloads sent outside the Earth's atmosphere. Although our
discussion focuses on Gaia, the concepts and considerations developed here
about sources of chromatic errors and how to deal with them can apply to all
experiments wishing to reach astrometric accuracy down to the mas-level
and beyond.\\
The aspects evidenced by our study of chromaticity are of critical importance
for compiling detailed error budgets and deciding on actions for controlling
systematic errors. Also, the degree of detail in instrument and measurement
models required to keep chromaticity at the $\rm{\mu as}$-level can help
prevent (during design), monitor (throughout operations), and correct
(in data reduction) other possible sources of systematic errors asso\-cia\-ted
with an image profile and its temporal and spatial variations. A possible
method for chromaticity correction is described in \cite{Gai2005} (2005).\\
In Sect. 2, we recall the relevant basic principles of a global astrometric
mission like Gaia\footnote{For details and recent updates consult the web site
  http://www.rssd.esa.int/index.php?project=SA}. In Sect. 3, we introduce
chromaticity for an all-reflecting optical system through a simplified
analytical model, which is also utilized to prove the linear dependence of
chromaticity on wavelength.
% *****************************************************************
\begin{figure*}[t]
\centering
\includegraphics[width=190pt,angle=-90]{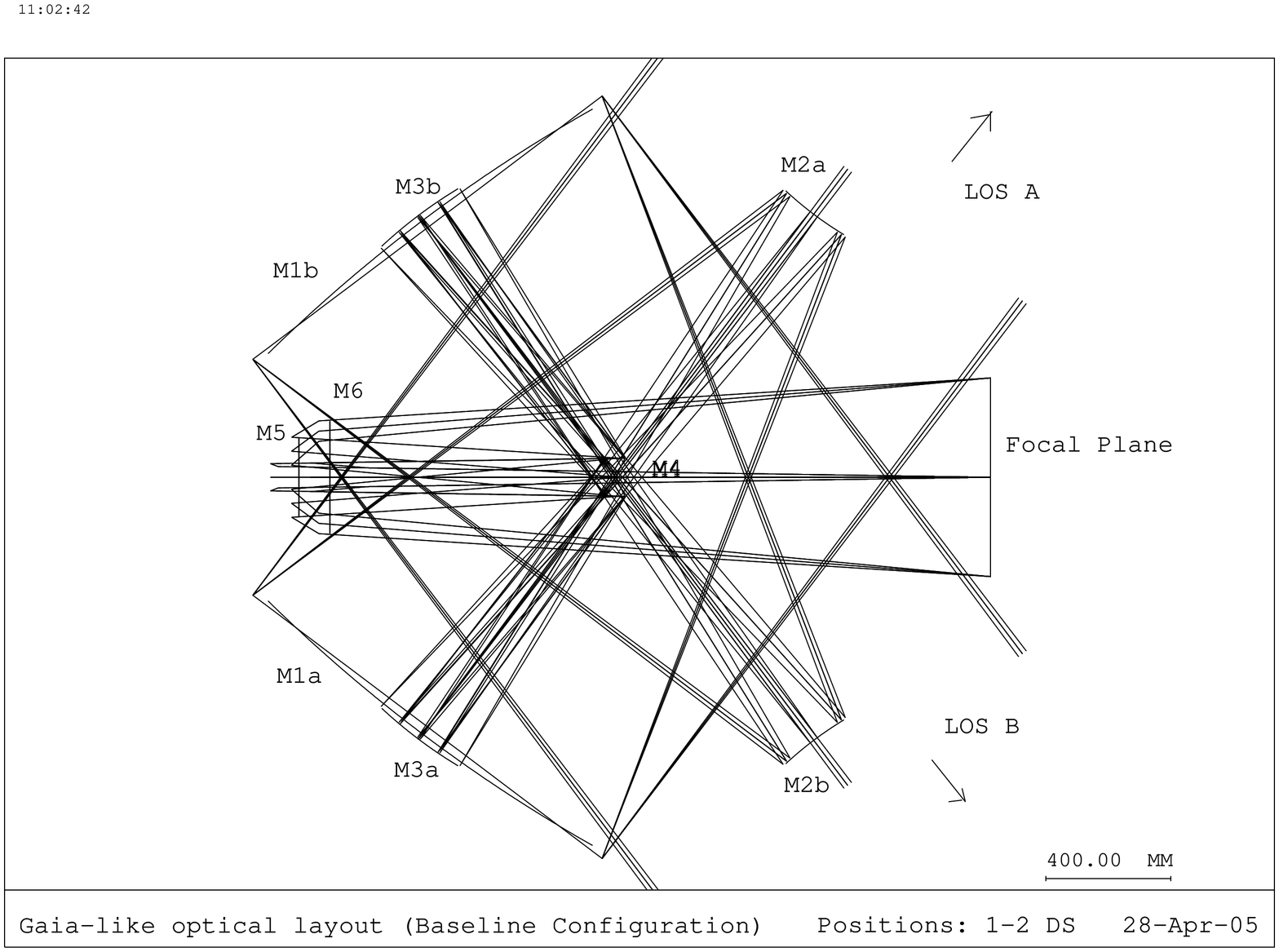}
\includegraphics[width=190pt,angle=-90]{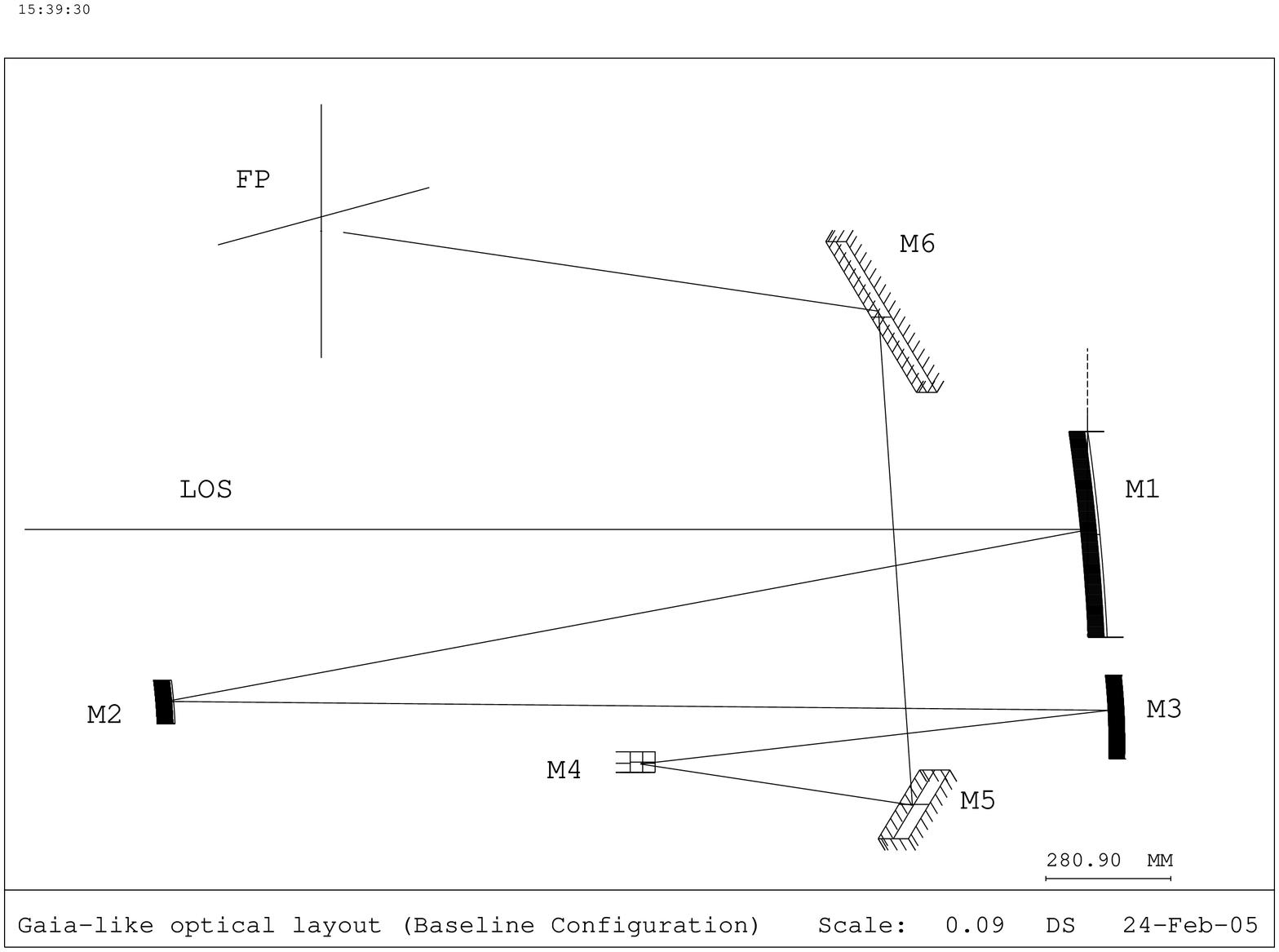}
\caption {A Gaia-like astrometric payload layout. Left: the set of
two telescopes combined at mirror M4 (planar drawing) Right:
individual telescope layout. LOS A and LOS B are the two lines of sight
implemented in this two-telescope configuration.}
\label{config}
\end{figure*}
% ***************************************************************** 
Section 4 presents the chromaticity analysis for a realistic representation of
the Gaia optical system. Section 5 provides prescriptions for minimizing
chromaticity in optical designs. In Sect. 6, we present the field-of-view
(FOV) distribution of chromaticity and me\-thods for its reduction. Finally,
conclusions are drawn in Sect.7.
% *****************************************************************
\section{Telescopes for global astrometry}
\label{section2}
Global astrometry is the direct measurement of angular position, parallax, and
proper motion, i.e. of those fundamental parameters that define, along with the
line-of-sight (LOS) velocity, the location and status of motion of every
object in space. The success of the Hipparcos mission established that the only
meaningful way to achieve global astrometry is from a space-born platform.
Suitable observing strategies and a sufficiently long operational life-time
would allow the whole sky to be covered repeatedly thus providing the means to
derive positions and their variations with time. 
As the result of renewed interest in  global astrometry by the astronomical
community worldwide, several missions are being considered by the major space
astronomy communities. In Japan studies are underway on the near-infrared
astrometry mission JASMINE (\cite{Jasmine}) for a possible launch in 2014. In
the United States, the NASA funded concept study for the possible development
of the Origins Billion Star Survey (OBSS)  is close to completion
(\cite{OBSS}), while teams at Jet Propulsion Laboratory (JPL) are well into
the development of SIM (Space Interferometry Mission), due for launch in 2011
(\cite{SIM}).\\
In Europe, following formal approval in 2000, the European Space Agency (ESA)
is developing its new astrometric mission Gaia, which is based on the same
two-FOV's scanning satellite concept implemented on Hipparcos, and is meant to
increase the measurement capabilities of its predecessor by about two orders
of magnitude.

Two different optical systems were considered during the study phase,
so that we will not know the optical configuration that will fly on Gaia before
final selection of the indu\-strial contractor (expected by early
2006). However, this is not a problem for the scope of this work as both
designs are based on all-reflective optics, and we can still refer to one
of them without loss of generality. We chose to refer to the configuration
known in Gaia jargon as the "Baseline Configuration"
(\cite{Perryman2}). It consists of the combination of two off-axis,
three-mirror monolithic telescopes mounted on a common optical bench, as
shown in Fig. 1 (left). The individual telescope has an effective focal length
(EFL) of 46.7m and a rectangular collecting area of $1.4 \times
0.5\ \rm{m^{2}}$, resulting in a central lobe size (analogous to the Airy
diameter for circular pupils) of $\sim 250 \times 700$ mas at the effective
wavelength $\lambda_{\rm{eff}}$ = 700 nm. We constructed a computer model of
the Baseline Configuration by utilizing the CODE V optical modelling package
(\cite{ORA}), and optimized the system over a FOV of
$0^\circ.66 \times 0^\circ.66$ by closely following the procedure described in
Loreggia et al. (2004). With reference to Fig. \ref{config} (right), the
optical train of each telescope consists of seven elements: three mirrors (M1,
M2, M3) with optical power, three flat mirrors (M4, M5, M6) for beam-folding,
and the focal plane (FP). The FOV covers the angular ranges $[-0^\circ.33, 0^\circ.33]$ in the along-scan direction and $[0^\circ.20, 0^\circ.86]$ in the across-scan direction;
the off-axis design is required to avoid vignetting. One of the possible
implementations investigated for the Gaia FP assembly calls for an array of 17
(along-scan) x 10 (across-scan) CCDs working in time delay integration (TDI)
mode (\cite{Saint-Pe}). Each along-scan strip of devices hosts: two chips for
the functions of target detection and confirmation, eleven CCDs for the
repeated astrometric measurement, four units for broadband photometry. The
rationale for photometry is both astrophysical (stellar classification) and
technical (chromaticity correction discussed below).\\ During each
transit, the image of a target on any of the eleven astrometric CCDs is binned
across scan, and centered along scan, providing independent one-dimensional
location estimates of comparable precision (\cite{Gai} and re\-fe\-ren\-ces therein). The composition of repeated along-scan measurements, taken for every
location on the sky with different orientations, allows the determination of
the bi-dimensional angular coordinates in a common reference system (``global
sphere reconstruction''). The evolution of the apparent source position during
the mission lifetime ($\sim 5$ years) yields absolute parallax and proper
motion.\\ Note that the relative orientation of the two telescopes, i.e. the
{\it base angle} (BA), is a key factor of the measurement process: the
on-sky separation between two stars entering the two different FOV's is
obtained from the measured difference in FP coordinates plus the BA;
therefore, configuration perturbations of a few {\it nm}, negligible as to the
optical response, are critical to astrometry as they can induce errors up to
several tens $\rm{\mu as}$.
Laser-based metrology lines are expected to monitor and (if necessary) to
control the BA stability that also sets stringent constraints on payload
thermal stability (\cite{Gardiol1}).
% ***********************************************************
\section {Chromaticity}
\subsection{ A basic analytical model }
\label{basic-model}
Chromaticity in all-reflective optical systems can be defined as the shift in
the FP location of the images of two sources of different spectral types
sharing the same direction in object space. Also, chromaticity is an intrinsic
characteristic of diffraction that cannot be neglected when the targeted
accuracy level is at the mas level and higher, as was the case for
Hipparcos and will be for Gaia.

We verified the presence of chromaticity is mainly related to wavelength;
sampling of the telescope point spread function (PSF) and dimensions of the
read-out region of the FP play a secondary role in the actual amount of
chromaticity displayed by the realization of any all-reflective optical
design.\\ Also, we will see that for a symmetric PSF there is no
chromaticity apart from possible residual effects due to sampling and other
noise sources (e.g. photon noise).  On the other hand, for an
asymmetric\footnote{In quite different optical configurations, the location
  of the peak of the diffraction pattern appears to be stable with respect to
  the spectral distribution of the celestial objects; this is consistent with
  the fact that, in case of reasonably limited aberrations, this maximum is
  coincident with the chief ray of geometric optics, which by definition is
  insensitive to wavelength variations. Therefore, the diffraction peak would
  appear as a good candidate for the PSF "photo-center" (location), if only
  the PSF were sampled with infinite resolution. In practice, it is impossible
  to measure it directly because of the finite pixel size.} PSF, most of the
chromatic effect, ${\Delta}$, can be modelled for every location estimators
tested as a linear function of the change in effective wavelength
$\delta\lambda_{\rm{eff}}$, characterizing the spectral difference of the
sources, through a coefficient $g(x,y)$ that depends on the local shape of the PSF,
$f(x,y)$, i.e. on the contributions of the aberrations over the FOV:
% ***********************************************************
\begin{equation}
{\Delta}= g(x,y)\cdot \delta\lambda_{\rm{eff}}, \label{chrom}
\end{equation}
% ***********************************************************
where ({\it x},{\it y}) are linear coordinates on the FP.

We illustrate relation (\ref{chrom}) with the help of ad hoc analy\-tical model. We choose an asymmetric, one-dimensional bell-shaped curve,
in the form of a deformed Gaussian profile, i.e. of different width on the
opposite sides of its peak value (at $y=0$)\footnote{As we make reference to a
  Gaia-like system in this work, we consider the y axis for consistency with
  most of the literature on Gaia. There the y-axis is the axis of the
  one-dimensional location process and the axis along which the scanning
  occurs.}:
% ***********************************************************
\begin{equation}
f(y)= \frac{1}{\lambda\sigma_{0}\sqrt{2{\rm \pi}}} \cdot \exp \left[-
\frac{1}{2}\left(\frac{y}{\lambda\sigma}\right)^{2}\right],
\hspace{0.1in}\sigma = \sigma_{0}+\epsilon\frac{y}{|y|},
\label{analytical-chrom}
\end{equation}
% ***********************************************************
$\epsilon$ being the "asymmetry" factor ($\epsilon <<\sigma_0$). For
continuity of the PSF, we must set
% ***********************************************************
\begin{equation}
\sigma(0)=\sigma_{0} \hspace{0.25in};\hspace{0.25in} f(0)=
\frac{1}{\lambda\sigma_{0}\sqrt{2{\rm \pi}}} .
\end{equation}
% ***********************************************************
Note that the position of the maximum is the origin, independent of
$\lambda$.\\
For two different wavelengths $\lambda_{1}<\lambda_{2}$, the normalization
condition requires that the maximum value decreases as the curve width
increases. This agrees with the physical condition that in the diffraction
limit, the PSF of an
optical system increases in linear size with the wavelength. In particular,
the PSF is a function of the wavelength $\lambda$, the position on the FP $y$,
and of the characteristic linear dimension $D$ of the optical system, in a
well-defined form required for dimensional reasons:
% ***********************************************************
\begin{equation}
PSF(y,\lambda,D) = PSF\left(\frac{yD}{\lambda}\right).
\end{equation}
% ***********************************************************
This is the typical form of the argument of the diffraction pattern for the
analytical reference cases solved in the literature. Therefore, a change in
wavelength corresponds, for the PSF, to a scaling of the FP coordinates; any
parameter weighted with this distribution will be affected by such variation.\\
Let us evaluate the behavior of, e.g. the center of gravity (COG) estimator,
for the model in Eq. (\ref{analytical-chrom}) over the region of interest $[-a,a]$:
% ***********************************************************
\begin{eqnarray}
\langle y \rangle = \int_{-a}^{+a} yf(y)&{\rm d}y=
\frac{\lambda}{\sqrt{2{\rm \pi}}}\left[4\epsilon + \frac{(\sigma_0-\epsilon)^2}{\sigma_0}\cdot\exp{-\frac{1}{2}\left[\frac{a}{\lambda(\sigma_0-\epsilon)}\right]^2}\right.
\nonumber\\
&\left.-\frac{(\sigma_0+\epsilon)^2}{\sigma_0}\cdot\exp{-\frac{1}{2}\left[\frac{a}{\lambda(\sigma_0+\epsilon)}\right]^2}\right]\sim
\frac{\lambda4\epsilon}{\sqrt{2{\rm \pi}}}, 
\end{eqnarray}
% ***********************************************************
where the approximation is valid for $a > \sigma_0 \lambda$.
The COG represents one of the possible estimators of the
photo-center location. In the Gaia literature, for example, the photo-center
location of a realistic, digitized image is named ``centroid'' for each
estimator used. Nevertheless, in this paper we retain the expression COG for 
ease of the reader. 
% ***********************************************************
The photo-center is linearly dependent on the selected wavelength through an
asymmetry factor $\epsilon$, representing the difference between the two sides
of the PSF, and the chromaticity relation (\ref{chrom}) becomes:
% ***********************************************************
\begin{equation}
{\Delta}={\delta \langle y \rangle}= \frac{4\epsilon}{\sqrt{2{\rm \pi}}}\cdot
\delta\lambda_{\rm{eff}}  . \label{chrom1}
\end{equation}
% ***********************************************************
In a more general case, the parameter $\epsilon$ can be considered a slowly
varying function of the focal plane coordinates, i.e. $\epsilon(x,y)$, to
account for the variation of the optical response (aberration distribution)
over the FOV. Also, different location estimators would generate functional
forms different from that in Eq. (\ref{chrom1}). 
% ************************************************************
\subsection{Asymmetric PSF with coma}
\label{asy-psf-coma}
The results of the previous section can be verified, in a more general way,
from the Fraunhofer integral for a circular aperture of radius $a$:
% ************************************************************
\begin{equation}
U(u,v)={\rm C} \int_0^1\int_0^{2{\rm \pi}} \exp[{\rm i}(k\Phi - v\rho \cos(\theta
- \psi) + u\rho^{2}/2)] \rho {\rm d}\rho {\rm d}\theta  ,   \label{fraun}
\end{equation}
% ************************************************************
where C is a normalization constant, $(\rho,\theta,z)$ and $(r,\psi,z)$ are
the cylindrical coordinates in the exit pupil and focal plane (image) spaces,
respectively. Then $\Phi$ is the term of aberration, $0\leq \rho\leq1$, and the
dimensionless variables $u$ and $v$ are defined as:
% ************************************************************
\begin{equation}
u=\frac{2{\rm \pi}}{\lambda}\frac{a}{R}z \hspace{0.5in}
v=\frac{2{\rm \pi}}{\lambda}\frac{a}{R}r \hspace{0.5in} r=\sqrt{x^2+y^2},
\end{equation}
% ************************************************************
{\it R} being the radius of the Gaussian sphere (=EFL). Following the treatment
in Born and Wolf (1980), we assume working with no piston ($u=0$) and with the
tangential coma as the only aberration term; this type of aberration is
expressed in the Nijboer-Zernike representation as
$\Phi={\rm A}_{\rm c}(3\rho^3-2\rho)\cos(\theta)$ (${\rm A}_{\rm c}$ is a coefficient
that quantifies the amount of coma in wavelengths). In this case, it can be
shown, that the solution of Eq. (\ref{fraun}) takes the form of a series:
% ************************************************************
\[
U(u,v,\psi)=U(0,v,\psi)=C[U_0(0,v,\psi)+
\]
\begin{equation}
\hspace{0.2in}+({\rm i}\alpha_{031})U_1(0,v,\psi)+(i\alpha_{031})^2U_2(0,v,\psi)+...],
\label{psfcoma}
\end{equation}
% ************************************************************
where $\alpha_{031}= 2{\rm \pi} {\rm A}_{\rm c} /\lambda$ and the functions $U_{\rm i}$ are
given as a combinations of Bessel functions of different orders, $J_{\rm k}$:
% ************************************************************
\[
U_0(0,v,\psi)= \frac{2J_1(v)}{v}; \hspace{0.3in}
U_1(0,v,\psi)={\rm i}\cos(\psi)\frac{2J_4(v)}{v} ;
\]
\[
U_2(0,v,\psi)=\frac{1}{2v}\left[\frac{1}{4}J_1(v)-\frac{1}{20}J_3(v)+\frac{1}{4}J_5(v)\right]
-
\]\
\begin{equation}
\hspace{0.3in} -
\frac{1}{2v}\left\{\frac{9}{20}J_7(v)+\cos(2\psi)\left[\frac{2}{5}J_3(v)+\frac{3}{5}J_7(v)\right]\right\}.
\label{uterms}
\end{equation}
% ************************************************************
\begin{figure}[t]
\centering \hspace{-0.2in}
\includegraphics[width=220pt]{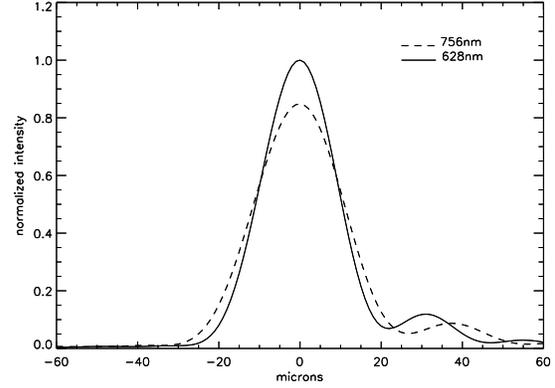}
\caption{The LSF obtained from the Fraunhofer integral with the coma
term $\Phi$ (see text).} \label{LSF}
\end{figure}
% ************************************************************

We used Eqs.(\ref{psfcoma}) and (\ref{uterms}) and the value\footnote{In
  practice, the optical system would be optimized to reduce
  coma. Consequently, the actual variation of the COG would, of course, be
  smaller.} ${\rm A}_{\rm c}=0.3$ (in wavelengths) for the coma coefficient to compute 
  the PSF. Hereafter we use the one-dimensional version, the Line Spread
  Function (LSF), obtained from the PSF by across-scan integration. The resulting curve is shown in
  Fig.\ref{LSF} for the two reference spectral types considered, B3V and M8V
  in the case of a Gaia-like configuration with $a=0.7\ {\rm m}$ and
  $R={\rm EFL}=46.7\ {\rm m}$. Note that the intrinsic effective wavelengths of B3V and M8V stars are about
  150 nm and 1300 nm, respectively. After taking realistic
  telescope transmission and detector response (\cite{CCD_QE}) into account, the effective
  wavelengths become $\lambda_{{\rm B3V}}=628$ nm and $\lambda_{{\rm
  M8V}}=756$ nm.
Using the discrete representation of the COG:
% ***********************************************************
\begin{equation}
y_{{\rm COG}} = \frac{1}{F} \sum y_{n}f_{n}, \label{COG1}
\end{equation}
% ***********************************************************
with $f_n$ the signal intensity recorded by the pixel in $y_n$, and $F =
\sum f_n$ the total intensity, we verified that the variation of the COG
position is linear with wavelength in the range of interest
following the relation $\delta y_{{\rm COG}}=-1.0195\lambda + 0.6415$, in agreement to the result derived in Sect. \ref{basic-model}.
The calculation was carried on by making reference to a situation using LSF
sampling and focal plane detectors, very similar to one of the options
considered for Gaia (\cite{short}): single CCD units of about 50 mm in the
scan direction, and pixels of $10\ {\rm \mu m}$ in the same direction. The
difference between the COG locations of the LSFs of our two reference star types amounts to $|{\rm COG_{B3V}}-{\rm COG_{M8V}}|= 0.121 {\rm mas}$.\\
%%%%%%%%%%%%%%%%%%%%%%%%%%%%%%%%%%%%%%%%%%%%%%%%%%%%%%%%%%%%%%%%%%%%%%
With our PSF modelling and measuring code we were able to investigate the
effects of limited detector dimensions (read-out region coincident with the
detector dimension) and sampling resolution (pixel size). We went on to extend
the detector dimension and to shrink the sampling step by two orders of
magnitude (CCD units of 50 cm on a side with $0.1\ {\rm \mu m}$ pixels) and
repeated the COG location measurements, which yielded: $|{\rm COG_{B3V}}-{\rm
  COG_{M8V}}|=0.122\ {\rm mas}$. This is a variation of only $1\%$ compared to
the value above.\\ 
%%%%%%%%%%%%%%%%%%%%%%%%%%%%%%%%%%%%%%%%%%%%%%%%%%%%%%%%%%%%%%%%%%%%%%%%
These experiments suggest, within the limits probed (the dimension of the
"unlikely" CCD simulated is very large but not infinite, and its pixels small
but not infinitesimal), that chromaticity is an intrinsic property of
all-reflective optical systems and that it can be approximated by a linear
function of source effective wavelength. Therefore whatever the detector
geometry and its spatial extension, different spectral type stars, set in the
same {\it nominal} position in object space, do not have the same {\it
  estimated} position on the FP, even with the simplest possible algorithm;
i.e. the same location on the sky is not uniquely mapped on the focal plane.\\
Chromaticity, a color-dependent position variation, is an important
contribution to the systematic error and must be removed to preserve the
desired mission accuracy.  
%****************************************************************
\section {Chromatic astrometric error}
\begin{figure}[t]
\centering
\includegraphics[width=250pt]{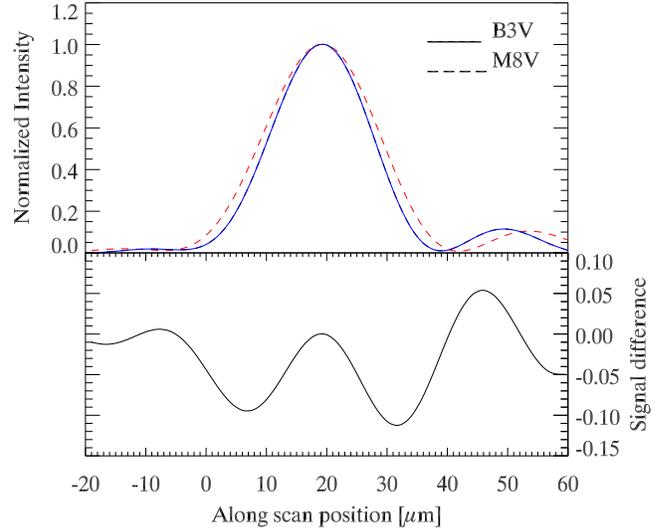}
\caption{Upper: PSF for B3V (solid line) and M8V (dashed line) stars in a
  representative off-axis field position; lower: PSF difference. }
\label{B3M8}
\end{figure}

Using our CODE V re\-pre\-sen\-ta\-tion of the Gaia Baseline Configuration, we could
derive Gaia-like images of the two B3V and M8V sources (all aberrations terms
are now taken into account), for any representative off-axis field
position. The sources are always intended to be in the same location in object
space, i.e. along the same direction on the sky. An example of the
one-dimensional optical signals are shown in the upper panel of
Fig. \ref{B3M8}, normalized to the peak value; it is apparent that the M8V
image (dashed line) is not just a geometrically scaled version of the B3V case
(solid line). The different spectral content weighs the design aberrations
(i.e. the residual aberrations after final optimization of the optical
design), and the diffraction image is also affected by profile variations, in
addition to the magnification associated to the intrinsic diffraction factor
$\lambda / D$. The signal difference is shown in Fig. \ref{B3M8} (lower): it
is not symmetric, as would be the case for perfect scaling, and reaches 15\%
of the initial peak value. For consistency with the treatment of the previous
section, the coordinate system is defined so that the COG location of an
ideal, ``non-aberrated'', PSF has abscissa equal to 0; the large common mode
displacement (the PSF's do not peak at 0) is mostly due to classical
distortion. It can be shown that different algorithms are affected by similar,
although not coincident, chromatic effects. Nevertheless, because of the
effective image profile of the Gaia telescope, the difference in performance
is relatively small (few percent), so that we continue to prefer the simplicity
of the COG algorithm throughout this work.\\
The application of the COG method for centering the Gaia-like PSF's of the two
reference stars yielded 20.112 ${\rm \mu m}$ for the blue star and 20.055
${\rm \mu m}$ for the red star, respectively, a difference of 57 nm. At the
optical scale of Gaia, about 4''/mm, this results in more than 200 ${\rm \mu
  as}$, i.e. comparable with the random location error for stars brighter than
${\rm V = 15 ~ mag}$.
% *****************************************************************
% *****************************************************************
\subsection{Monochromatic and polychromatic PSF's}
\label{mono-poli}
It is possible to simplify the formalism used in sect. \ref{asy-psf-coma} by
recalling that the PSF can be derived by taking the square modulus of the
complex amplitude response function (ARF), which in turn is derived as the
Fourier transform of the generalized pupil function $P_{\rm g}$
(\cite{Goodman}), i.e.

% ***********************************************************
\begin{equation}
ARF(x,y; \lambda) = q(\lambda) \times F\{P_{{\rm g}}(\rho,\theta)\} \ ,
\label{ARF}
\end{equation}
% ***********************************************************
where $(x,y)$ are coordinates on the focal plane and $(\rho, \theta)$
are taken on the pupil.

The Fourier integral in Eq. (\ref{ARF}) is performed over the rectangular
pupil of the Gaia telescope, where the factor $q(\lambda)$ is constant.
The pupil function $P_{\rm g}$ is expressed in terms of the wavefront
error (WFE) function, $W(\rho,\theta)$, via the relation
% ***********************************************************
\begin{equation}
P_{{\rm g}}(\rho,\theta; \lambda) = \exp \left[{\rm i} \frac{2{\rm
      \pi}}{\lambda}W(\rho,\theta) \right] \ . \label{pupil}
\end{equation}
% ***********************************************************
Note that Eq. (\ref{pupil}) clearly shows how at longer wavelengths the effect
of the WFE is reduced, as the phase contribution is a smaller fraction of the
period.\\
The expression for the monochromatic PSF, $PSF_{\rm m}$, is therefore
% ***********************************************************
\begin{equation}
PSF_{\rm m}(x,y,\lambda) = \left| ARF(x,y;\lambda) \right|^2 .
\end{equation}
% ***********************************************************
The local polychromatic $PSF(x,y)$ for a star with a given spectral
distribution is
% ***********************************************************
\begin{equation}
PSF(x,y) = \int_{\lambda_1}^{\lambda_2} \omega(\lambda) \times PSF_m(x,y,\lambda) \
{\rm d}\lambda \ . \label{PSF}
\end{equation}
% ***********************************************************
The weight function $\omega(\lambda)$ combines the source spectral
distribution, the instrument transmission, and the quantum efficiency of the
detector; then the polychromatic PSF can be normalized to the expected average
number of detected photons.\\
CODE V can compute the WFE distribution associated to a given field position
for a selected optical configuration; it is thus possible to build the PSF for
any desired source, using Eqs. (\ref{ARF}) -- (\ref{PSF}). We use a
numeric implementation of this model (\cite{Busonero}) to provide the
description of the imaging performance of the Gaia telescope with all the
aberration terms given by CODE V. Additional contributions describing the
realistic detector response and the effect of TDI observations can be
included, as well as the across-scan binning used for Gaia, to build the
recorded signal. For any source of known spectrum, it is possible to derive
its effective wavelength, the detected signal, and the chromaticity with respect to the selected reference spectral
type using the above expressions. In this way, the measurement of Gaia can, in principle, be made {\it
  achromatic}, as the position estimate is no longer function of the source
spectral distribution. This correction requires both color information and
knowledge of the local instrument response.
%**************************************************
\begin{table*}[t]
\caption{Effects of individual Standard Zernike terms vs. the
non-aberrated (ideal) case.}
\label{aberration1}
\centering
\begin{tabular}{cccccc}
\hline
  \hline
  Term & Standard &  RMS WFE  &    Image RMS          &  COG displacement & Chromaticity\\
       & Zernike  &   [nm]    & width increase [$\%$] &     [mas]         &   [$\mu$as] \\
  \hline
  no. aber. & / & 0 & 0 & 0 & 0 \\
  1 & 1 & 0 & 0 & 0 & 0 \\
  2 & $\rho \cos(\theta)$ & 11.37 & 0 & 0 & 0 \\
  3 & $\rho \sin(\theta)$ &  32.44 & 0.1346 & 34.29 & 97.95 \\
  4 & $\rho^2 \cos(2\theta)$ & 13.02 & 1.263 & 0.030 & -2.22 \\
  5 & $2\rho^2-1$ & 26.03 & 4.927 & 0.119 & -9.27 \\
  6 & $\rho^2 \sin(2\theta)$ & 26.48 & 0.3199 & 21.35 & -359.7 \\
  7 & $\rho^3 \cos(3\theta)$ & 11.76 & 1.881 & 0.045 & -3.651 \\
  8 & $(3\rho^3-2\rho)\cos(\theta)$ & 12.51 & 1.881 & 0.045 & -3.691 \\
  9 & $(3\rho^3-2\rho)\sin(\theta)$ & 32.49 & 1.455 & -33.38 & 779.6 \\
  10 & $\rho^3\sin(3\theta)$ & 13.61 & 0.829 & 14.34 & -118.3 \\
  11 & $\rho^4\cos(4\theta)$ & 8.928 & 1.079 & 0.028 & -1.042 \\
  12 & $(4\rho^4-3\rho^2)\cos(2\theta)$ & 23.15 & 5.026 & 0.144 & 11.64 \\
  13 & $(6\rho^4-6\rho^2+1)$ & 34.21 & 9.965 & 0.262 & 20.22 \\
  14 & $(4\rho^4-3\rho^2)\sin(2\theta)$ & 38.69 & 2.083 & -36.64 & 1019 \\
  15 & $\rho^4\sin(4\theta)$ & 6.423 & -0.008 & 0.125 & -652 \\
  16 & $\rho^5\cos(5\theta)$ & 5.597 & 0.3312 & 0.097 & 0.6501 \\
  17 & $(5\rho^5-4\rho^3)\cos(3\theta)$ & 246.53 & 7.817 & 0.234 & 26.46 \\
  18 & $(10\rho^5-12\rho^3+3\rho)\cos(\theta)$ & 25.99 & 6.186 & 0.175 & 14.92 \\
  19 & $(10\rho^5-12\rho^3+3\rho)\sin(\theta)$ & 20.61 & 3.329 & 2.229 & -2142 \\
  20 & $(5\rho^5-4\rho^3)\sin(3\theta)$ & 26.35 & 4.571 & -18.66 & 1372 \\
  21 & $\rho^5\sin(5\theta)$ & 5.376 & -0.169 & -1.346 & -393.4 \\
  \hline
\end{tabular}
\end{table*}
%**************************************************
\subsection{Identification of the critical WFE terms}
The source of chromatic errors is the WFE, which is present in any nominal
optical design with a finite FOV; manufacturing and alignment may
only aggravate the problem. Besides, not every component of WFE contributes to
chromaticity: below, we analyze the effect of individual aberrations and the
related symmetry properties.\\
We then investigate how to minimize, by design, the chromatic errors in any
given field position, analyzing the effect of partial or total suppression of
selected aberration terms. Finally, we proceed to evaluate the overall field
properties of chromaticity, verifying the possibility of compensating for it by
taking measurements over a complete crossing of the focal plane, a possibility
offered by scanning instruments like Gaia.\\
In optical engineering, typical expansions of the WFE are in terms of Zernike and Fringe Zernike polynomials,
orthogonal, and normalized functions, if mapped on a circular pupil
(\cite{Born}). In our analysis we have used expansions with 21 Zernike terms;
the set of standard Zernike functions are listed in Table \ref{aberration1}
(for Fringe Zernike equivalents see \cite{Born}). Note that the representation
in Zernike polynomials is not optimal for non-circular pupils and more
convenient expansions have been investigated (\cite{Gardiol2}).\\ 
In the following, we show how to identify the main individual contributors to chromaticity in the case of a
realistic representation of the Gaia telescope regardless of the choice of the
Zernike set, and to search for possible
correlations between chromaticity and parameters like RMS WFE and image RMS
width.\footnote{The root mean square wavefront error, RMS WFE, is simply the
  square root of the quantity $\int WFE^2(\rho , \theta)$ taken over the
  pupil. The image RMS width for, e.g. the $y$ axis on the focal plane is
  computed as $ \sqrt{\int (y- < y >)^2 L(y) dy} $, where $ L(y) = \int
  PSF(x,y) dx $ is the line spread function.}\\
%***********************************************
Chromaticity is again evaluated by taking the difference of the COG positions
obtained for the adopted reference spectral types B3V and M8V. We used
both monochromatic PSFs at the reference wavelengths associated with the two
spectral types and the full polychromatic representation of Eq. (\ref{PSF}),
assuming the two stars behave like blackbodies. The latter description is more
representative from the astrophysical standpoint, whereas the former is
computationally much simpler. Preliminary experiments proved that the two
methods can be considered equivalent when, as is the case here, the primary
goal of the investigation is to evaluate of the properties of chromaticity
and not its precise calculation.\\
The image RMS width is derived from the monochromatic PSF at $\lambda = 700\
{\rm nm}$, roughly representative of solar type objects and sort of midway
between B3V and M8V types. The ima\-ge COG displacement is always referred to
the ideal, non-aberrated case, which has an image RMS width of $8.6425\ {\rm
  \mu m}$. Each Zernike term is individually evaluated, with the coefficient set
to 0.1 (i.e. the small aberrations regime is assumed): then the  WFE is built
from the selected term, and the PSF computed accordingly to the above
model. The results for the first 21 terms are listed in Table
\ref{aberration1}.\\
%***************************************************************
\begin{figure}[t]
\centering
\includegraphics[width=250pt]{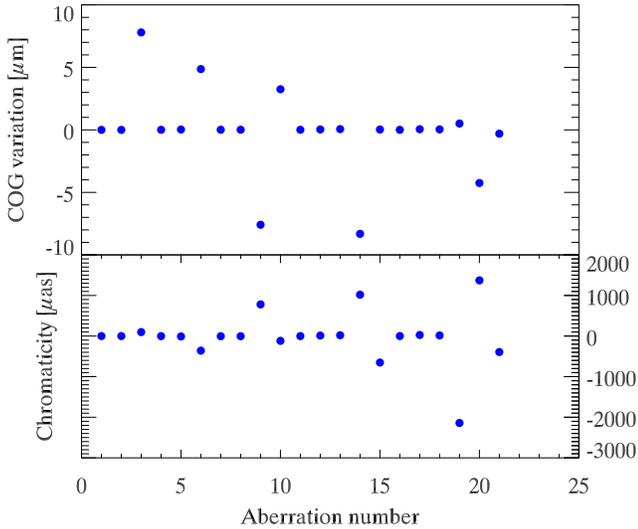}
\caption{Image COG variation (upper panel) and chromaticity
variation (lower panel) vs. Standard Zernike aberration terms.} \label{IMAG}
\end{figure}
%********************************************************************
In Fig. \ref{IMAG}, we show the impact of each individual aberration on the COG
(i.e. an effect corresponding to classical distortion) and on
chromaticity. Some terms strongly contribute to both, but in general there is
little correlation. Similarly, we found that there is no simple relationship
between RMS WFE, or image RMS width (both including the contribution of all
aberrations) and chromaticity.\\
The complex relationship between WFE, image RMS width, and chromaticity can be
understood in terms of symmetry, since the aberrations are in general
bidimensional, whereas the parameters relevant to the Gaia measurement are
mostly one-dimensional, i.e. referred to the scan (y-axis) direction. Thus,
specific aberration terms may contribute to the WFE significantly, to
across-scan image width and across-scan (x-axis) centroid displacement, with
little impact on the location, noise and chromaticity, due to the across-scan
binning of the images (see Sect. \ref{section2}).\\
There is a trend toward increasing image RMS width with RMS WFE, but this is not a
strict relation, and several cases of large WFE and small image width
degradation are seen. This could be associated to aberrations inducing
significant image degradation in the across scan direction only. The COG, in some cases, is affected by a large displacement associated to low
chromaticity values. Thus, the images are translated with respect to the
non-aberrated position by an amount that does not depend on the spectral distribution
of the source. The case of classical distortion fits this description.

From Fig. \ref{IMAG}, we see that significant contributions to chromaticity,
i.e. $\geq 100\ {\rm \mu as}$, come only from aberrations (no. 3, 6, etc.) with
a specific functional form: they are all the odd (sinusoidal) functions of the
angular coordinate, so that they have odd symmetry on the pupil plane. All of
the even (cosinusoidal) terms do not provide net chromaticity; values below
$\sim 30\ {\rm \mu as}$ are the result of limited precision in the calculation
scheme that is implemented.\\
%****************************************************************
A similar analysis was carried on with the Fringe Zernike
polynomials. Again, the critical terms for chromaticity we identified to be
those associated to the odd parity portion of the WFE, i.e. the sinusoidal
terms, as for the Standard Zernike expansion.  Therefore, in the case of Gaia,
the terms that should be minimized, by design, manufacturing, and alignment
are those with odd parity (anti-symmetric) relative to the across-scan ($x$)
axis.
% ***********************************************************
\section{Compensation by aberration selection}
\label{section5}
\begin{figure}%[h]
\hspace{-0.2in}
\centering
\includegraphics[width=250pt]{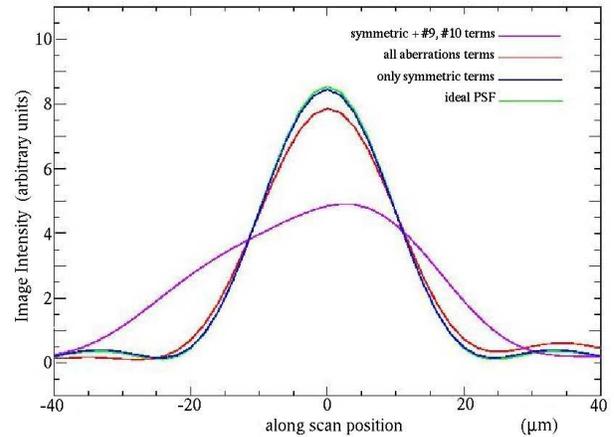}
\caption { Modifications of the PSF profile due to the different weight of
  aberration terms for field position F4 of the Gaia-like astrometric optics considered here.}
\label{aberr_select}
\end{figure}

We now turn our attention to verifying chromaticity with many
aberration terms present. The analysis refers to FP position F4 (see Table
\ref{ch_field}) with coordinates $0^\circ.20$ (along-scan) $\times
0^\circ.33$ (across-scan), which is affected by a significant value of
chromaticity \footnote{ As for the previous section, chromaticity is evaluated
  as the difference between the monochromatic PSF at 628 nm and 756 nm,
  respectively. The chromaticity derived with the fully polychromatic model is
  about a factor 2 larger.}. Six cases are considered:
\begin{enumerate}
\item non-aberrated (ideal) PSF,
\item nominal case (from optical design),
\item removal of all symmetric terms,
\item removal of all anti-symmetric terms,
\item removal of anti-symme\-tric terms AND
scaling of symmetric terms to equivalent RMS WFE,
\item subset of anti-symmetric terms 9 and 10.
\end{enumerate}
The corresponding values of RMS WFE, image RMS width, and chromaticity are
listed in Table \ref{sixcases}. The resulting PSF's (monochromatic at the
reference wavelength of 700 nm) for four of the cases above are shown in
Fig. \ref{aberr_select}.
%**************************************************************
\begin{table*}[t]
\caption {RMS WFE, image RMS width (for the two reference spectral types),
and chromaticity at different field angle locations over the focal plane.}
\label{ch_field}
\centering
\begin{tabular}{ccccccc}
\hline
  \hline
   & X position & Y position & RMS WFE & Image RMS width & Image RMS width & Chromaticity \\
   & [degrees]  & [degrees]  & [nm]    &  for a B3V star [mas] & for a M8V star [mas]
   & [${\rm \mu as}$] \\
   \hline
  F1 & 0.53 & 0.00 & 22.03 & 39.27 & 42.85 & 0.00 \\
  F2 & 0.86 & 0.33 & 208.73 & 41.80 & 43.89 & -3071.23 \\
  F3 & 0.86 & -0.33 & 208.73 & 41.81 & 43.89 & 3070.16 \\
  F4 & 0.20 & 0.33 & 39.77 & 41.93 & 44.92 & 2077.02 \\
  F5 & 0.20 & -0.33 & 39.77 & 41.93 & 44.92 & -2077.02 \\
  F6 & 0.70 & 0.17 & 24.32 & 38.51 & 42.54 & 124.94 \\
  F7 & 0.70 & -0.17 & 24.32 & 38.51 & 42.54 & -124.94 \\
  F8 & 0.36 & 0.17 & 28.82 & 40.30 & 43.34 & 2545.20 \\
  F9 & 0.36 & -0.17 & 28.83 & 40.30 & 43.34 & -2545.20 \\
  F10 & 0.86 & 0.00 & 71.09 & 38.95 & 42.76 & 0.00 \\
  F11 & 0.53 & -0.33 & 41.81 & 40.07 & 43.77 & -389.84 \\
  F12 & 0.53 & 0.33 & 41.81 & 40.07 & 43.77 & 389.84 \\
  F13 & 0.20 & 0.00 & 18.59 & 39.28 & 43.02 & 0.00\\
\hline
\end{tabular}
\end{table*}
% **********************************************************
The nominal case provides an image RMS width that is reasonably close to the
diffraction limit in spite of non negligible RMS WFE (40 nm, i.e. $\lambda /
15$ at $\lambda = 600\ {\rm nm}$); the chromaticity is about 1 mas.

Suppression of the symmetric aberrations provides some improvement to the RMS
WFE. The variation of the image width is marginal and the initial chromaticity is mostly
retained. Conversely, removing the anti-symmetric aberrations completely (case
4), we achieve significant improvement, as in the previous case, on the RMS
WFE, but not nearly as much on the image RMS width; chromaticity, however, is
reduced to zero, according to expectations. The RMS sum of WFE in cases 3
and 4 restores the nominal value: the mutual orthogonality of symmetric and
anti-symmetric function sets is preserved, even if this is no longer true for
the individual functions within each set. Even when the symmetric aberrations
are scaled to restore the initial WFE level of 40 nm (case 5), the
chromaticity is still zero.\\
Besides, when a random subset of anti-symmetric aberrations (9 and 10) is
retained, together with the symmetric terms, with the nominal coefficients
(case 6), the result has a value that is comparable to the chromaticity in the
nominal case (case 2) but with the opposite sign. Futhermore it leads to a
dramatic degradation of both WFE and image quality.
% ***********************************************************
\begin{table}
\caption{Changes in RMS WFE, image RMS width, and chromaticity for
the 6 selected cases discussed in sect. \ref{section5}.}
\label{sixcases}
\centering
\begin{tabular}{cccc}
\hline
  \hline
  Case & RMS WFE & Image RMS & Chromaticity \\
       &  [nm]   &   width [mas]        &    [mas] \\
   \hline
  1 & 0.00 & 39.26 & 0.000 \\       
  2 & 40.17 & 42.51 & 1.025 \\      
  3 & 33.94 & 41.75 & 1.011 \\      

  4 & 21.50 & 40.14 & 0.000 \\      
  5 & 40.17 & 42.23 & 0.000 \\      

  6 & 102.67 & 360.72 & -0.799 \\       
  \hline
\end{tabular}
\end{table}
%****************************************************************
This shows that the optimization procedure in the ray tracing code  actually
achieves some partial compensation among different aberrations, providing
some, although limited, benefit to chromaticity. On the other hand, standard
optical design optimization procedures are based on improving the general
image quality parameters such as WFE and spot diagrams. Therefore, it is possible
to introduce custom merit functions in the optimisation procedure, which can 
include computation of the chromaticity or of the critical anti-symmetric
contribution to WFE, with some averaging rule over the field. This does not
necessarily modify the actual optical configuration in any signi\-fi\-cant
way, since it is always convenient to start {\it after} standard optimization,
but chromatic aberrations can be further reduced at the expense of the others
and, possibly, of a small increase in overall WFE, which is
acceptable in many cases.
%**************************************************************
% **********************************************************
\section{Field distribution of chromaticity}
\begin{figure}
\centering
\includegraphics[width=250pt]{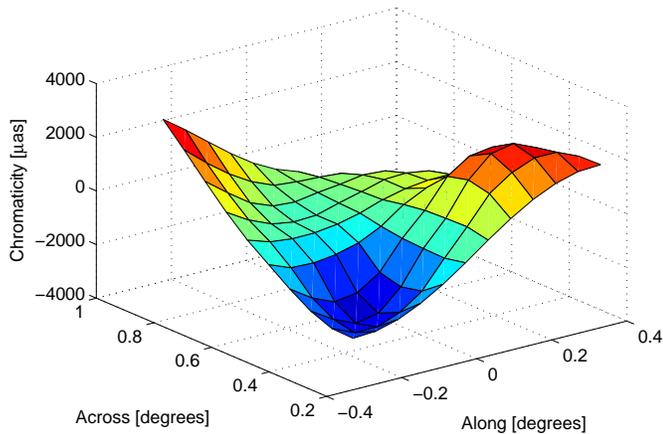}
\caption {Chromaticity distribution over the astrometric field in the nominal
  configuration of the Gaia-like astrometric payload.}
\label{field_ch}
\end{figure}

The map of chromaticity over the FOV was derived from the optical
representation of the baseline configuration of the astro\-me\-tric
payload. We consi\-de\-red only the nominal design aberrations; manufacturing
and alignment errors are not included yet. The result can therefore be
considered as a {\it best case}.\\
The analysis used monochromatic PSFs in the set of positions
listed in Table \ref{ch_field} that also lists the associated RMS WFE in
nanometers (nm), the image RMS width for the two cases considered (B3V and M8V
sources), and the chromaticity. The values were interpolated to cover the
field with $0^{\circ}.02$ resolution, and the resutling surface is shown in
Fig. \ref{field_ch}. The distribution is antisymmetric with respect to the $y$
(across scan) axis, due to the intrinsic symmetry of the optical
configuration: the mean chromaticity value is $-0.05\ {\rm \mu as}$,  with RMS
value $1.19\ {\rm mas}$ and peak values exceeding $\pm 3\ {\rm mas}$,
i.e. three orders of magnitude larger than the Gaia measurement goal.
% ***********************************************************
\subsection{Transit level compensation}
The symmetry of the chromaticity distribution can be exploited to reduce the
overall contribution to a set of measurements, at transit-level, in spite of
comparably high local values. All targets detected by Gaia are
observed in TDI mode on the whole FP, so that the measurements
are performed in opposite positions with respect to the symmetry axis; by
composition of the photo-center values from exposures in symmetric positions
along scan, the re\-si\-dual chromaticity cumulated over a transit drops to
values quite close to zero.\\
% ***********************************************************
%************************************************************
The average value of the transit chromaticity over the astrometric field is
$-0.10\ {\rm \mu as}$, with RMS value $0.21\ {\rm \mu as}$. Therefore,
chromaticity compensation over the transit appears quite effective in the
nominal configuration.
%************************************************************
\subsection{Partial compensation on misaligned systems}
\begin{figure}
\centering
\includegraphics[width=220pt]{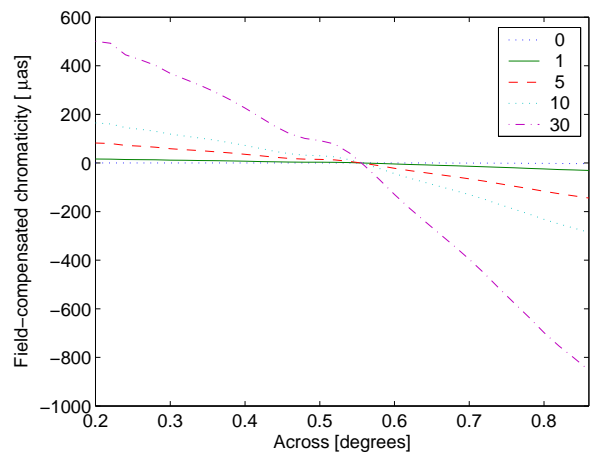}
\caption {Transit level chromaticity vs. across scan field position in the
  nominal case and for increasing alignment error, (in arcsec).} 
\label{transit_ch}
\end{figure}

The real configuration used in flight will not retain the design symmetry, due
to manufacturing, integration, and re-alignment errors. In order to model the
effect, the chromaticity map derived for the nominal configuration is simply
shifted by an amount corresponding to the selected offset ranging from
one arcsecond to 0.5 arcminute (a fairly large alignment error). The local
chromaticity value is mostly preserved, but the transit level combination,
which is very close to zero in the nominal (symmetric) case, is affected by a
degradation increasing with the error, as shown in Fig. \ref{transit_ch}. The statistics of transit-level chromaticity across the FP is shown in Table
\ref{transtab}. Column 1 lists the offset applied (zero corresponds to the
nominal case); in cols. 2 and 3 we report average and standard
deviation of transit-level chromaticity, respectively, as computed across the
field.\\ 
The transit-level chromaticity remains very close to zero in an across-scan
position of about $0^{\circ}.55$, which appears to be a chromatic-free section
of the field; the residual has opposite signs on either side of this position. 
This is due to the structure of the local chromaticity distribution
(Fig. \ref{field_ch}), with alternate signs in each quadrant. The result
is that stars of a given spectral type in different regions of the field have
nearly opposite residual chromaticity. This aspect may be exploited for
further reduction of the residual chromaticity in the data reduction
phase. Upon definition of a threshold of acceptable residual chromaticity, at
transit level, it is possible to provide a specification for alignment,
referred to both initial telescope integration, and in-orbit re-alignment.
% ***********************************************************
\begin{table}
\caption{Transit-averaged chromaticity as a function of
re-alignment error.}
\label{transtab}
\centering
\begin{tabular}{ccc}
\hline
\hline
Offset   & Chromaticity Mean & Chromaticity RMS \\
(arcsec) &    (${\rm \mu as}$)     &   (${\rm \mu as}$)    \\
\hline
0  &  -0.403 &   0.8218 \\
1  &  -2.599 &  14.1906 \\
5  & -11.277 &  68.3465 \\
10 & -21.892 & 136.5464 \\
30 & -61.998 & 414.4432 \\
\hline
\end{tabular}
\end{table}
% ************************************************************
% ***********************************************************
\subsection{Optical engineering aspects}
Symmetry considerations can be applied to the contribution from the individual optical components, in particular concerning their figuring error in the 
manufacturing phase, which can be split in components relevant or not to 
overall chromaticity. 
Also, alignment aspects, above referred to the primary mirror, can be applied 
with appropriate modifications to perturbation of other components. 
Manufacturing and
alignment errors inducing an anti-symmetric residual WFE on a telescope mirror
result in increased chromaticity. The on-flight configuration will be affected
by manufacturing and alignment errors, which can also be described in terms of
symmetric and anti-symmetric aberrations. If the design is optimized in terms
of low chromaticity, small perturbations due to manufacturing and alignment
are likely to induce a comparably low increase in the chromatic errors, due to
larger tolerances. With the configuration considered as representative for
Gaia, the relevant values of residual transit chromaticity are on the order of
a few
arcseconds at the level of the primary mirror, which is challenging for
standard optical techniques. However, it is not necessary to perform
re-alignment of {\it each} optical component: as for many astronomical
instruments, when the errors are not too large, a global re-optimization can
be achieved by adjusting a small number of degrees of freedom, often
localized on a single component (e.g. the secondary mirror). The correction
requirements in terms of stroke and resolution are within the range of the
actuators considered for Gaia; the main limiting factor is the diagnostics
capability, which in any case will benefit from accurate analysis of the
image properties over the field.\\
Quantitative analysis and optimization may be performed on specific optical
configurations, including manufacturing and alignment aspects based on
realistic WFE data from manufacturers. The resulting minimization of the
instrumental chromaticity is an improvement to the overall systematic error
budget, which is desirable because correction procedures based on the
science data are necessarily limited, and in some ways they subtract
information.
\section{Conclusions}
This paper has discussed the issue of chromaticity in all-reflective optical systems. In
spite of common belief, avoiding refractive components is not sufficient for
achieving an achromatic instrument response, due to basic diffraction
considerations. In the scenario of future space astrometric missions, the
impact of systematic errors two or three orders of magnitude larger than the
measurement goal is of fundamental importance.\\
Specific analyses have been performed with reference to the baseline
configuration of the Gaia astrometric payload, but the assumptions,
principles, and conclusions of our discussion can be applied to any high
accuracy astrometric instrument. The chromatic error is defined as the
difference in image photo-center location at different wavelengths, and the
exact value depends on the selected measurement process, but the effect is
unavoidable.\\
Independently from the selected WFE expansion, the terms relevant to
chromaticity are those associated with anti-symmetry of the PSF in the FP and
with anti-symmetric WFE contributions on the pupil plane. Symmetric terms only
contribute to the astrometric noise by increasing the effective image width
in the measurement direction. As the relation between RMS WFE and chromaticity
is complex, the specification of only the RMS WFE is not a sufficient
requirement for controlling chromaticity from the optical manufacturing
standpoint.\\
The first prescription to optical manufacturers is to suppress or at least
minimize the anti-symmetric terms. However, it is not necessary to set all
chromatic terms to zero: an appropriate combination is still able to provide
some local balancing. Standard optical design optimization techniques are able
to provide reasonable results by applying the conventional image quality merit
functions; optimal results on chromaticity require  definition of {\it ad hoc}
criteria, as much as on tolerancing.  The distribution of chromaticity over the
field inherits some symmetry properties from the optical system; deviations
from symmetry are induced e.g. by manufacturing and alignment errors on each
optical component.\\ 
In case of repeated measurements in different parts of the field, some
chromaticity compensation is achieved in the data combination, depending on
the symmetry of both instrument and measurement schemes. In the case of Gaia,
each object observed by a symmetric (i.e. correctly re-aligned) telescope
along a full transit provides a set of astrometric measurements affected by
opposite chromatic errors in symmetric positions; transit-level composition is
therefore likely to remove a large fraction of the local chromaticity. Any
residual chromaticity must be removed in the science data processing after the
best implementation of the astrometric payload, to minimize the initial
systematic error. This requires spectral information for each source {\it and}
a good knowledge of the detailed instrument response (\cite{Gai2005} 2005). 
% ***********************************************************
\begin{acknowledgements}
We acknowledge the contribution of Dr. D. Carollo to initial investigations of
the subject of chromaticity. The ORA team helped in the detailed
operation and best usage of their CodeV package. Our activity on Gaia was
partially supported by the Italian Space Agency, under research contracts ASI
ARS 96-77 and ASI ARS 98-92. M.G.L. acknowledges the support of the Space
Telescope Science Institute through the Institute's Visitor Program for 2005.
\end{acknowledgements}
% ***********************************************************

% ***********************************************************
\end{document}